\documentclass[aps,prl,showpacs,twocolumn,floatfix]{revtex4}

\usepackage{graphicx}
\usepackage{amsmath}
\usepackage{amssymb}
\usepackage{bm}
\usepackage{dcolumn}

\begin{document}

\title{Magnetic Phase Diagram of FeAs based superconductors}
\author{Zhicheng Zhong, Qinfang Zhang, P. X. Xu and Paul J. Kelly}
\affiliation{Faculty of Science and Technology and MESA$^+$
Institute for Nanotechnology, University of Twente, P.O. Box 217,
7500 AE Enschede, The Netherlands}

\date{\today }

\begin{abstract}
The recently discovered high-temperature superconductivity in doped
quaternary iron oxypnictides correlates experimentally with a magnetic
instability. We have used first-principles calculations to determine a
magnetic phase diagram of ReO$_{1-\delta}$FeAs (Re=La--Dy) as a
function of the doping $\delta$, of the FeAs in-plane lattice constant
$a$, and of the distance between the Fe and As planes, that is
qualitatively consistent with recent experimental findings on the
doping, internal (chemical) and external pressure dependence. The
existence of a tricritical point (TCP) in the phase diagram suggests
new ways of enhancing $T_c$.

\end{abstract}

\pacs{
}

\maketitle


The recent discovery of  superconductivity in electron-doped
La[O$_{1-x}$F$_x$]FeAs with a critical temperature ($T_c$) of 26~K
\cite{Kamihara:jacs08} has stimulated a massive experimental
\cite{Takahashi:nat08,Hunte:nat08,Chen:nat08,delaCruz:nat08,Chen:prl08,%
Ren:cpl08,Ren:epl08b,Ren:epl08a,Bos:cc08,Dong:epl08,Nakai:jpsj08,
Kohama:prb08,Drew:prl08,Takabayashi:jacs08} and theoretical effort
\cite{Haule:prl08,Cao:prb08,Ma:prb08,Singh:prl08,Boeri:prl08,%
Yildirim:prl08,Yin:prl08,Ishibashi:jpsj08,Mazin:prb08,Giovannetti:physb08,%
Mazin:prl08,Kuroki:prl08,Fang:cm08} to find other, higher $T_c$
materials in this completely new family of iron-pnictide
superconductors. The undoped parent compound LaOFeAs is a poor metal
with an ordered antiferromagnetic (AFM) ground state
\cite{delaCruz:nat08,Dong:epl08} but with increasing F doping
\cite{Nakai:jpsj08} the magnetic ordering is suppressed and
superconductivity emerges. This strongly suggests that magnetic
fluctuations in the iron layers close to the quantum critical point
(QCP) play a fundamental role in the superconducting pairing mechanism.
Recent experimental
\cite{Nakai:jpsj08,Kohama:prb08,Drew:prl08,Liu:prl08} and theoretical
\cite{Giovannetti:physb08,Mazin:prl08,Kuroki:prl08,Fang:cm08} results
also suggest that spin fluctuations in the vicinity of the QCP mediate
the superconductivity as in the cuprates, heavy fermion materials, or
ruthenates \cite{Mathur:nat98,Moriya:rpp03,Monthoux:sci07}. In view of
this, so far empirical, correlation between superconductivity and QCPs,
it is important to understand how the magnetic (in)stability depends on
structural and chemical parameters that are accessible to experiment.
That is the subject of this paper.

Superconducting ReOFeAs crystallizes in a tetragonal layered structure
with $P4/nmm$ symmetry and consists of layers of covalently bonded FeAs
alternating with layers of more ionically bonded ReO. With eight atoms
(two formula units) in the unit cell, it can be described using only
two internal structural parameters in addition to the lattice
parameters $a$ and $c$ \cite{Quebe:jac00}. One of the internal
parameters, $d_{\rm Fe-As}$, describes the separation between planes of
Fe and As; the other, $d_{\rm Re-O}$, between planes of Re and O. The
excess electron from the ReO layer is donated to the FeAs layer which
is metallic with a number of partly filled bands of mainly Fe $3d$
character intersecting the Fermi energy
\cite{Lebegue:prb07,Singh:prl08}. In addition to the four structural
parameters, we define a doping $\delta$ in terms of deviations from
this ideal stoichiometry per Fe atom. Electron and hole doping can be
achieved, for example, by partially replacing oxygen with fluorine
\cite{Kamihara:jacs08} or trivalent La with divalent Sr
\cite{Wen:epl08}.

Since mapping out the magnetic phase diagram in five dimensions is
impossible, we need to identify a smaller number of key independent
variables. Clearly the doping $\delta$ is one. Although the interaction
between the ReO and FeAs layers is by no means negligible, it is widely
accepted that the superconducting properties of ReOFeAs emerge from the
FeAs layers. The main role of the ReO layers is to determine the
lattice parameters \cite{Quebe:jac00} and to contribute doping
electrons while the magnetism and superconductivity are associated with
the ``active'' FeAs layers so we fix $d_{\rm Re-O}$ and $c$ at their
calculated equilibrium values for undoped LaOFeAs. The in-plane lattice
constant $a$ and the separation between the Fe and As layers $d_{\rm
Fe-As}$ \cite{Yildirim:prl08,Yin:prl08,Ishibashi:jpsj08,Mazin:prb08}
are the other two key parameters we identify to construct magnetic
phase diagrams in the $(\delta, a=a_{eq}, d_{\rm Fe-As})$ and
$(\delta=0.25, a,d_{\rm Fe-As})$ planes of this parameter space. We do
this by calculating the energy $E(\delta, a, d_{\rm Fe-As})$ for
non-magnetic (NM), checkerboard AFM (C-AFM) and stripe AFM (S-AFM)
orderings, going considerably beyond existing attempts to determine the
QCPs of FeAs-based materials as a function of doping or pressure
\cite{Giovannetti:physb08,Fang:cm08}.

The total energy calculations are carried out within the framework of
density functional theory (DFT) using a spin polarized generalized
gradient approximation (SGGA) for the exchange-correlation potential
(PW91 functional). The electronic ground state is calculated by solving
the Kohn-Sham equations self-consistently with the projected augmented
wave method (PAW) \cite{Blochl:prb94b} and a cut off energy of 500 eV
for the plane wave basis as implemented in the Vienna \emph{ab initio}
simulation package (VASP) \cite{Kresse:prb96}. The C-AFM and S-AFM
states are described in a $\sqrt{2} \times \sqrt{2} \times 1$
tetragonal structure. Though an orthorhombic distortion is observed
\cite{delaCruz:nat08,Yildirim:prl08} for the S-AFM ordered state in the
LaFeAsO parent compounds, it is suppressed by doping. Since we are
mainly interested in locating the phase boundaries, it can be
neglected. The Brillouin zone integrations are performed with the
improved tetrahedron method \cite{Blochl:prb94a} with a sampling grid
of 6 $\times$ 6 $\times$ 4 k-points plus $\Gamma$ point. When using
pseudopotentials, it has been shown that it is necessary to treat the
$3p$ states of Fe as valence states to reproduce all-electron results
\cite{Mazin:prb08}. Doping was modelled by adding or substracting
electrons and compensating their charges with a homogeneous fixed
background charge \cite{Kresse:prb96}. This approximation was checked
explicitly using supercell calculations in which a fraction of the
oxygen atoms were replaced with fluorine, and proved not to be
critical.

\begin{figure}[t]
\includegraphics[scale=0.5, angle=0]{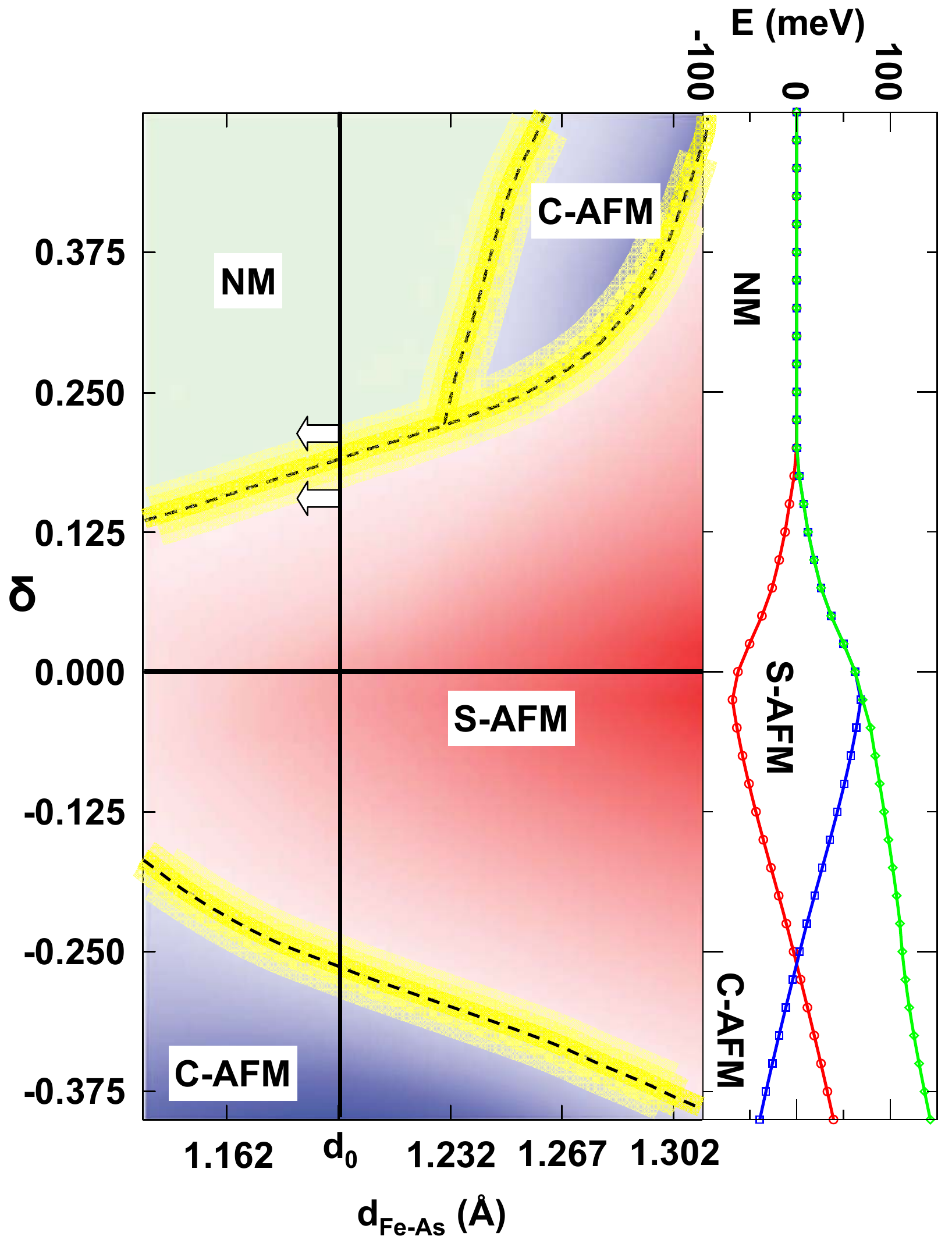}
\caption{Phase diagram of LaOFeAs as a function of doping $\delta$ and
the distance between the Fe and As planes along the $c$ axis, $d_{\rm
Fe-As}$. Blue, red and green represent respectively checkerboard AFM
(C-AFM), stripe AFM (S-AFM) and nonmagnetic (NM) ordering. In the AFM
region, more stable magnetic ordering is described by darker color.
$d_0$ is the calculated equilibrium value of $d_{\rm Fe-As}$ for the
undoped material. On the right hand side we show the energies per Fe
atom of C-AFM, S-AFM and NM ordered states relative to their average
value as a function of doping for the vertical line $d_{\rm
Fe-As}=d_0$. The white arrows describe the pressure dependence of over-
and under-doped LaOFeAs.}
\label{dopingphase}
\end{figure}

We begin by determining the ground state of LaOFeAs as a function of
$\delta$ and $d_{\rm Fe-As}$. First of all the energy is minimized with
respect to the internal structural parameters $d_{\rm Fe-As}$ and
$d_{\rm Re-O}$ for the undoped parent compound LaOFeAs using the
experimental $a$ and $c$ \cite{fn:ac}. This optimized structure is then
frozen and the total energy calculated as a function of $d_{\rm Fe-As}$
and $\delta$ for NM, C-AFM and S-AFM ordering. The phase diagram
obtained from these energies is shown in Fig.~\ref{dopingphase} where
positive and negative $\delta$ correspond to electron and hole doping,
respectively. We propose that this phase diagram describes
qualitatively all FeAs based materials since substituting different
elements in the Re-O layers only changes the effective doping $\delta$
and the lattice parameters $a$ and $c$. The effect of changing $c$, for
example by external uniaxial pressure, is given in our phase diagram in
terms of its effect on $d_{\rm Fe-As}$; changing the Re position has
otherwise little effect. The effect of changing $a$ will be discussed
below. Because our phase diagram is calculated as a function of two
variables, the boundaries separating regions with different ground
states, shown as dashed lines in Fig.~\ref{dopingphase}, are in general
quantum critical {\em lines} (QCL). The uncertainty in the location of
these quantum critical lines resulting from neglecting the effect of
differential relaxation specific to the magnetic ordering and doping,
and because the SGGA is not exact, is indicated by the broader yellow
regions.

The most notable feature of Fig.~\ref{dopingphase} is the asymmetry of
electron and hole doping. The undoped material exhibits S-AFM ordering
for all values of $d_{\rm Fe-As}$ (horizontal line, $\delta=0$) and
S-AFM ordering is stabilized by increasing $d_{\rm Fe-As}$. On doping
LaOFeAs with electrons (vertical line, $d=d_0$), the S-AFM ordering is
weakened and a transition to NM ordering is observed at $\delta \sim
0.18$. The suppression of magnetic ordering by electron doping is
consistent with previous first-principles calculations
\cite{Giovannetti:physb08,Xu:cm08}. We can use this phase diagram to
explain the dependence of $T_c$ on external pressure. Experimentally it
has been found that $T_c$ increases with external pressure for $x <
0.14$ and decreases for $x > 0.14$ for Sm[O$_{1-x}$F$_x$]FeAs
\cite{Takabayashi:jacs08}. This behavior indicates the existence of a
QCP around $x \sim 0.14$. According to Fig.~\ref{dopingphase} the
over-doped arsenide does not order magnetically while the under-doped
($x < 0.14$) material favours S-AFM order. External pressure will
reduce $d_{\rm Fe-As}$ (white arrows) but its effect on $T_c$ depends
on the doping. For under-doped Sm[O$_{1-x}$F$_x$]FeAs it destabilizes
the S-AFM ordering and pushes the system towards the QCL making it more
susceptible to spin fluctuations. In agreement with observations
\cite{Takabayashi:jacs08}, we expect this to enhance $T_c$.  However,
the over-doped system is NM and pressure drives it away from the QCL
leading to lower values of $T_c$, as observed. When the structure is
doped with holes, a transition to C-AFM ordering occurs. We will
discuss below how the asymmetry of hole and electron doping can be
understood in terms of the density of states (DOS) of the parent
compound close to the Fermi level. Our phase diagram makes a clear
prediction for the pressure dependence of $T_c$ of hole doped iron
arsenides, namely that it should be precisely the same as the
electron-doped case electron-doped case.

\begin{figure}[t]
\includegraphics[scale=0.34, angle=0]{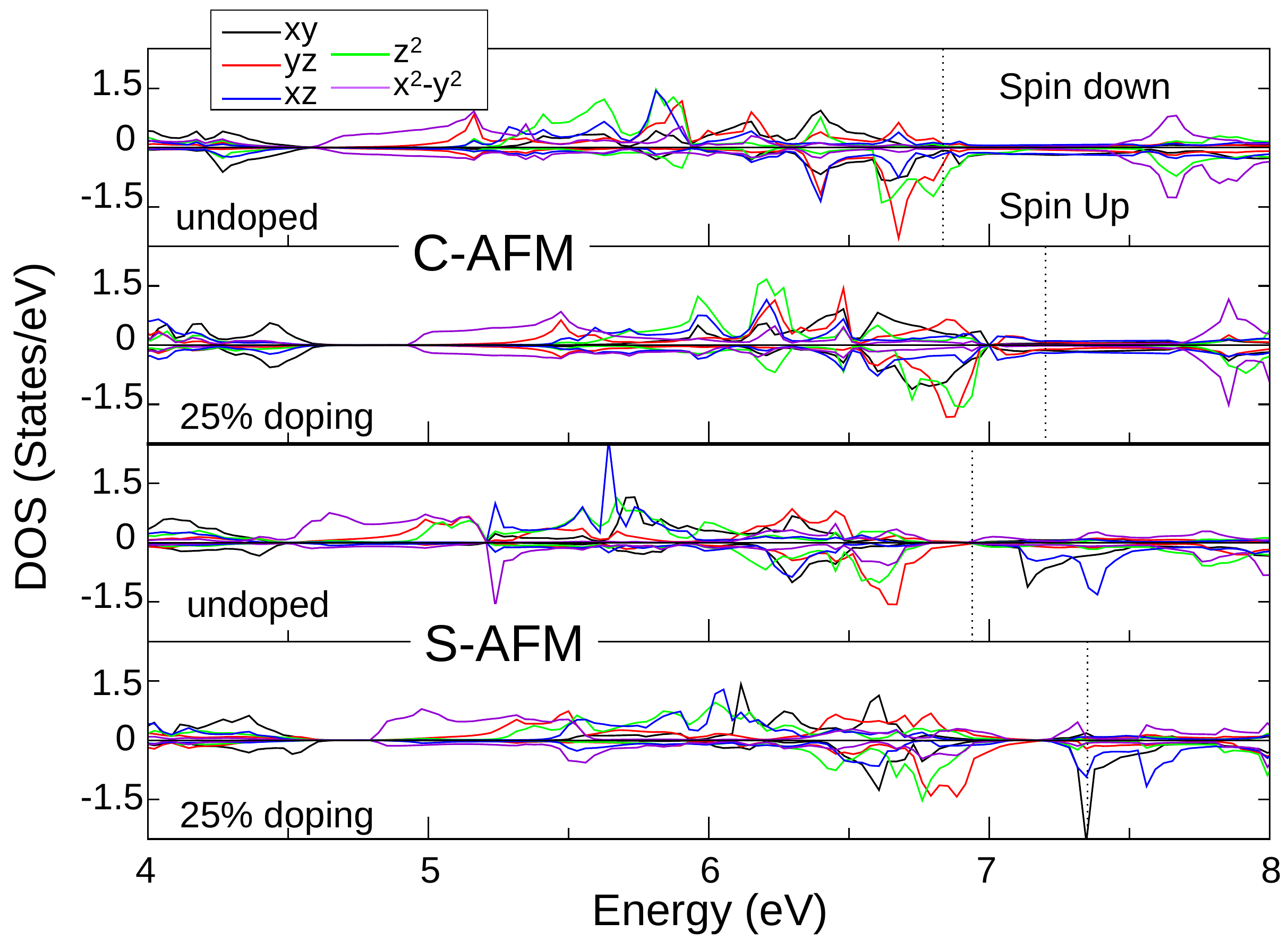}
\caption{Projected densities of states calculated for C-AFM ordered
undoped and 25\%-doped LaOFeAs; for S-AFM Fe ordered undoped and
25\%-doped LaOFeAs. The orbital characters are color-coded. The Fermi
energy is shown by the vertical dotted line.}
\label{dos}
\end{figure}

A second notable feature of Fig.~\ref{dopingphase} is the existence of
a narrow region between NM and S-AFM states where the ground state is
C-AFM for large $d_{\rm Fe-As}$ and high electron doping. The line
separating NM and C-AFM states meets the line separating C-AFM and
S-AFM states for an electron doping of $\delta \simeq 22\%$ and $d_{\rm
Fe-As} \simeq 1.03 d_0$ in a tricritical point (TCP) where NM, S-AFM,
and C-AFM states coexist. For electron doping $\delta > 22\%$,
increasing $d_{\rm Fe-As}$ generates a sequence of NM $\rightarrow$
C-AFM $\rightarrow$ S-AFM groundstates. To understand why this C-AFM
region exists, we examine the Fe projected DOS shown in Fig.~\ref{dos}
for C-AFM and S-AFM ordering for undoped and 25\% electron doping. For
S-AFM ordering, the Fermi energy in the undoped case is situated in a
pseudogap for both spin channels. On doping with electrons, the
pseudogap has to be crossed before a high density of (mainly
minority-spin) states peak with $d_{xy}$ and $d_{x^2-y^2}$ character
can be populated. This is energetically unfavourable. For C-AFM
ordering, no such pseudogap exists and the doping electrons can be
accomodated in states close to the undoped Fermi level while the large
value of $d_{\rm Fe-As}$ stabilizes the AFM states. It is the
competition between these two mechanisms that results in the existence
of a narrow region of C-AFM state.

In the context of heavy fermion and cuprate superconductivity, it has
been suggested that spin fluctuations close to QCPs can give rise to
the attractive interaction between carriers needed to mediate
superconductivity \cite{Mathur:nat98,Moriya:rpp03,Monthoux:sci07}. The
coexistence of NM, C-AFM, and S-AFM states at the TCP ($\delta \simeq
22\% ; d_{\rm Fe-As}\simeq 1.03 d_0$) suggests that spin fluctuations
there may be stronger and more interesting. We therefore propose
searching for higher values of $T_c$ in the vicinity of the TCP. This
requires simultaneous achievement of large electron doping and negative
pressures by suitable chemical substitutions.

%
\begin{figure}[t]
\includegraphics[scale=0.33, angle=0]{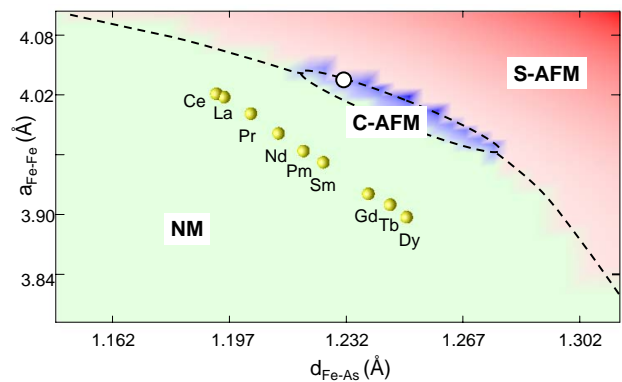}
\caption{Phase diagram of La[O$_{0.75}$F$_{0.25}$]FeAs as a function of
lattice parameter $a$ and inter-layer distance between Fe and As plane
along $c$ axis ($d_{\rm Fe-As}$). The structure parameters obtained by
geometry optimization for 25\% doped Re compounds are plotted as yellow
spheres. The color coding is the same as in Fig.~\ref{dopingphase}. The
white circle indicates the position of the tricritical point for a
doping of $\delta=0.22$. } \label{adphase}
\end{figure}

Chemical pressure can be exerted by replacing La with other rare earth
atoms, the in-plane lattice constant $a$ and unit cell volume
decreasing monotonically with decreasing Re size in the sequence La,
Ce, Pr, Nd, Sm, Gd \cite{Quebe:jac00}. The effect of these
substitutions has been studied experimentally resulting in the
observation of a maximum value of $T_c = 55$K for SmOFeAs
\cite{Ren:cpl08,Ren:epl08b,Ren:epl08a,Bos:cc08}. These substitutions
change not only the lattice constants but also the internal parameter
$d_{\rm Fe-As}$. To study how sensitively the QCPs depend on $a$, we
construct a phase diagram in the $(\delta=0.25, a, d_{\rm Fe-As})$
plane, Fig.~\ref{adphase}. To achieve this high electron doping, we
replace one out of four O atoms by a F atom in a $\sqrt{2} \times
\sqrt{2} \times1$ supercell in order to be able to more accurately
describe distortions induced by the high concentration of the smaller F
ion. The results agree well with those obtained on doping with the
jellium model used to construct Fig.~\ref{dopingphase}.

The main features of this phase diagram are (i) that the effect of
increasing $a$ is similar to that of increasing $d_{\rm Fe-As}$: it
stabilizes AFM ordering, and (ii) that the values of these structure
parameters are not optimal. A small C-AFM ``island'' appears at the
boundary between the NM and S-AFM ``seas'' in Fig.~\ref{adphase},
consistent with Fig.~\ref{dopingphase}. This C-AFM island represents an
area where spin fluctuations are expected to be large. For a doping
$\delta \sim 0.22$ corresponding to the TCP, this island shrinks to a
point located at $d_{\rm Fe-As}=1.23\AA$, $a=4.04\AA$, indicated in the
figure by the white circle. We optimize the geometries of
Re[O$_{0.75}$F$_{0.25}$]FeAs for Re =Ce, Pr, Nd, Pm, Sm, Gd, Tb, and Dy
by energy minimization and plot the resulting structure parameters as
yellow spheres in the figure \cite{fn:REpp}. Assuming that $T_c$ is
proportional to proximity to the TCP, we see that the Nd, Pm and Sm
compounds should have the highest $T_c$s in agreement with observation
\cite{Ren:cpl08}. The experimental data for ReO$_{1-\delta}$FeAs
(Re=La, Ce, Pr, Nd, Sm) \cite{Ren:epl08b} have been interpreted as
evidence for a lattice contraction induced enhancement of $T_c$. We
note that replacing La with heavier Re elements is accompanied not only
by a reduction of $a$ but also by an increase in $d_{\rm Fe-As}$. From
the phase diagram of Fig.~\ref{dopingphase}, we attribute the higher
$T_c$ primarily to the inecrease in $d_{\rm Fe-As}$. We expect that the
highest $T_c$ will occur for doped NdOFeAs and SmOFeAs and that
replacing Nd or Sm with heavier Re elements \cite{Bos:cc08} will not
lead to further improvement.

The phase diagram of Fig.~\ref{dopingphase} suggests that the optimal
electron doping for FeAs based superconductors is around 22\%. Electron
doping can be reasonably well controlled experimentally by replacing O
with F or by introducing oxygen vacancies. However
Figs.~\ref{dopingphase} and \ref{adphase} together indicate that
currently achieved values of $a$ and $d_{\rm Fe-As}$ are smaller than
the optimal values and should be increased to achieve higher values of
$T_c$. One way of doing this would be to replace some As with the
larger isovalent elements Sb or Bi or with Te.

In conclusion, we propose that the electronic structures and magnetic
properties of FeAs-based superconductors such as LaOFeAs,
BaFe$_2$As$_2$ and LiFeAs are dominated by the effective doping and
structural details of the FeAs layers. In a QCP scenario, the
superconductivity is related to spin fluctuations making it important
to understand the magnetic phase diagram as a function of
experimentally accessible parameters. Currently available experimental
data can be understood using the theoretically obtained phase diagrams
we have presented giving us confidence that they can then be used to
search for higher values of $T_c$.

\emph{Acknowledgments:} This work is part of the research program of
the ``Stichting voor Fundamenteel Onderzoek der Materie'' (FOM) and the
use of supercomputer facilities was sponsored by the ``Stichting
Nationale Computer Faciliteiten'' (NCF), both financially supported by
the ``Nederlandse Organisatie voor Wetenschappelijk Onderzoek'' (NWO).
It is also supported by EC Contract No. IST-033749 "DynaMax." The
authors wish to thank S. Kumar and G. Brocks for useful discussions.


\begin{thebibliography}{41}
\bibitem[{fn:({\natexlab{a}})}]{fn:ac}
\bibinfo{note}{The calculated SGGA values of $a$ and $c$ are 0.402 and 0.863,
  very close to the experimental values of 0.404 and 0.875 nm, respectively
  [\onlinecite{Quebe:jac00}].}
\bibitem[{fn:({\natexlab{b}})}]{fn:REpp}
\bibinfo{note}{Pseudopotentials for the Re elements Ce-Dy with the 4f electrons
  treated as core electrons and for La are calculated in slightly different
  ways so that the calculated structure parameters are not strictly
  comparable.}
\end{thebibliography}

\end{document}